\documentclass{ws-ijmpa}
\usepackage{amsmath}
\usepackage{amssymb}
\usepackage{cite}
\begin{document}

\markboth{\footnotesize B.F.L. WARD}
{MASSIVE ELEMENTARY PARTICLES AND BLACK HOLE PHYSICS IN RESUMMED QUANTUM GRAVITY}

\catchline{}{}{}{}{}
%
%%%%%%%%%%%%%%%%%%%%%%%%%%%%%%%%%%%%%%%%%%%%%%%%%%%%%%%%%%%%%%%%%%%%

\title{MASSIVE ELEMENTARY PARTICLES AND BLACK HOLE PHYSICS IN RESUMMED QUANTUM GRAVITY}

\author{\footnotesize B.F.L. WARD}

\address{Department of Physics, Baylor University, One Bear Place \#97316\\
Waco, Texas 76798-7316,
USA
}

\maketitle

%\pub{Received (Day Month Year)}{Revised (Day Month Year)}

\begin{abstract}
We use exact results in a new approach to quantum gravity to 
discuss some issues in black hole physics.
\end{abstract}

\section{Introduction}
%Albert Einstein showed that Newton's law, one of the most basic laws in
%physics, is a special case of the solutions of the
%classical field equations of his general theory of relativity.
% Specifically,
%$g_{00}=1+2\Phi_N \Rightarrow \nabla^2\Phi_N=4\pi G_N\rho$
%from
%$R^{\alpha\gamma}-\frac{1}{2}g^{\alpha\gamma}R  =-8\pi G_N T^{\alpha\gamma}$, etc., where he have now introduced the familiar metric of space-time
%$g_{\mu\nu}$, the Newtonian potential $\Phi_N$, Newton's constant
%$G_N$, the mass density $\rho$, the contracted Riemann 
%tensor $R^{\alpha\gamma}$,
%and the appropriate energy momentum tensor $T^{\alpha\gamma}$.
There have been several successful tests of Einstein's general theory 
of relativity in classical physics~\cite{mtw,sw1,abs}.
Heisenberg and Schroedinger, following Bohr, formulated a quantum mechanics
that has explained, in 
the Standard Model(SM)~\cite{sm}, 
all established experimentally accessible 
quantum phenomena except the quantum treatment of Einstein's theory. Indeed,
even with tremendous progress in quantum field theory,
superstrings~\cite{gsw,jp}, loop quantum gravity~\cite{lqg}, etc.,
no satisfactory treatment of the
quantum mechanics of Einstein's theory is known to be correct 
phenomenologically.
Here, with an eye toward black hole physics,
we apply a new approach~\cite{bw1} to quantum gravitational phenomena, 
building on previous work by Feynman~\cite{f1,f2} to get a minimal
union of Bohr's and Einstein's ideas.
%basic idea quantum gravity is a point particle 
%quantum field theory
%and its apparent bad UV
%behavior is due to our naivete --
%nothing fundamental prevents
%the union of Bohr's and Einstein's ideas.

The approaches to the to the attendant bad UV behavior
have been summarized in Ref.~\cite{wein1}. 
%There are four approaches~\cite{wein1} to the attendant bad UV behavior 
%of quantum gravity (QG):
%extended theories of gravitation such as supersymmetric theories - superstrings
%and loop quantum gravity;
%resummation, a new version of which we discuss presently;
%composite gravitons; and,
%asymptotic safety -- fixed point theory, recently pursued with
%success in Refs.~\cite{laut,reuter2}.
Our approach, based on YFS methods~\cite{yfs,yfs1},
is a new version of the resummation approach~\cite{wein1} and
allows us to make contact
with both the extended theory~\cite{wein1} 
and the asymptotic safety~\cite{laut,reuter2}
approaches and to discuss issues in black hole physics, some of which
relate to Hawking~\cite{hawk} radiation.\par

%Our new approach 
%which we have introduced in Refs.~\cite{bw1}
%, resummed quantum gravity, is based on well-tested YFS~\cite{yfs,yfs1}
%methods. We first review
%Feynman's formulation of Einstein's theory in Sect. 2. We present
%resummed QG in Sect. 3. In Sect. 4 we discuss Newton's law.
%In Sect. 5 we discuss the black hole physics, some of which is
%related to Hawking
%radiation~\cite{hawk}.
%Section 6 contains some summary remarks.

\section{Review of Feynman's Formulation of Einstein's Theory}

In the SM there are many massive point particles.
Are they black holes in our new approach to quantum gravity?
To study this question, we follow Feynman, treat spin as
an inessential complication~\cite{mlg}, and 
%replace $L^{\cal G}_{SM}(x)$ in (\ref{lgwrld}) with
consider the simplest case for our question, that of gravity coupled to 
a ``free'' scalar
field, a ``free'' physical Higgs field, $\varphi(x)$, with a rest mass 
$m$ believed to be less than $400$ GeV and known to be greater than $114.4$ GeV with a
95\% CL~\cite{lewwg}. 
%We are then led to consider the representative 
%model~\cite{f1,f2} {\small
%\begin{equation}
%\begin{split}
%{\cal L}(x) &= -\frac{\sqrt{-g}}{2\kappa^2} R
%            + \frac{\sqrt{-g}}{2}\left(g^{\mu\nu}\partial_\mu\varphi\partial_\nu\varphi - m_o^2\varphi^2\right)\\
%            &= \frac{1}{2}{\big\{} h^{\mu\nu,\lambda}\bar h_{\mu\nu,\lambda} - 2\eta^{\mu\mu'}\eta^{\lambda\lambda'}
%\bar{h}_{\mu_\lambda,\lambda'}\eta^{\sigma\sigma'}\\
%&\bar{h}_{\mu'\sigma,\sigma'}{\big\}}
%          + \frac{1}{2}{\big\{}\varphi_{,\mu}\varphi^{,\mu}-m_o^2\varphi^2 {\big\}} \\
%&-\kappa {h}^{\mu\nu}{\big[}\overline{\varphi_{,\mu}\varphi_{,\nu}}+\frac{1}{2}m_o^2\varphi^2\eta_{\mu\nu}{\big{]}}\\
%            & \quad - \kappa^2 [ \frac{1}{2}h_{\lambda\rho}\bar{h}^{\rho\lambda}{\big{(}} \varphi_{,\mu}\varphi^{,\mu} - m_o^2\varphi^2 {\big{)}} \\
%&- 2\eta_{\rho\rho'}h^{\mu\rho}\bar{h}^{\rho'\nu}\varphi_{,\mu}\varphi_{,\nu}] + \cdots \\
%\end{split}  
%\label{eq1}
%\end{equation}}\noindent
%where $\varphi_{,\mu}\equiv \partial_\mu\varphi$ and 
%we have the metric
%$g_{\mu\nu}(x)=\eta_{\mu\nu}+2\kappa h_{\mu\nu}(x)$ with
%$\eta_{\mu\nu}={\text diag}\{1,-1,-1,-1\}$ and 
%%%%\item {\Color{Blue}$R$} is the curvature scalar. \\
%%%%{\Color{Black} $\kappa=\sqrt{8\pi {\Color{PineGreen}G_N}}$}
%$\bar y_{\mu\nu}\equiv \frac{1}{2}\left(y_{\mu\nu}+y_{\nu\mu}-\eta_{\mu\nu}{y_\rho}^\rho\right)$ for any tensor $y_{\mu\nu}$.
%%The Feynman rules for (\ref{eq1}) were
The Feynman rules for this theory were 
already worked-out by Feynman~\cite{f1,f2}.
%where we use his gauge, $\partial^\mu \bar h_{\nu\mu}=0$. 
On this view,
%\item {\Color{Brown}$\varphi$} is representative of matter at a high scale 
%{\Color{Brown}$\sim M_{GUT}=10^{16}$GeV}
quantum gravity is just another quantum field theory
where the metric now has quantum fluctuations as well.
For example, 
the one-loop corrections to the graviton propagator
due to matter loops is just given by the diagrams in Fig. 1.
\begin{figure}
\begin{center}
\epsfig{file=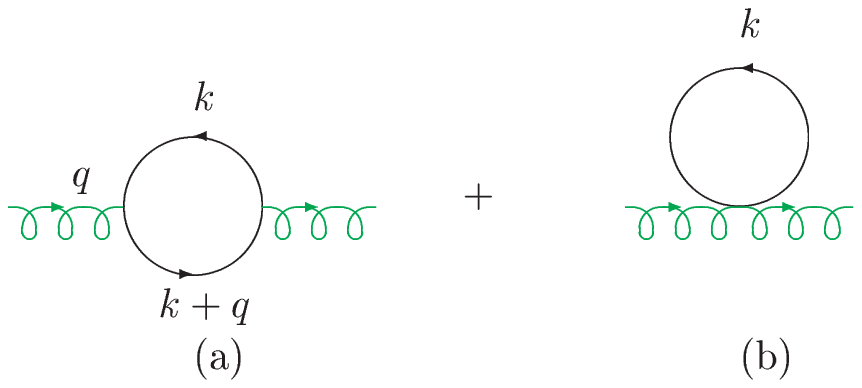,width=77mm,height=38mm}
\end{center}
\caption{\baselineskip=7mm     The scalar one-loop contribution to the
graviton propagator. $q$ is the 4-momentum of the graviton.}
\label{fig1}
\end{figure}
We return to these graphs shortly.

\section{Resummed Quantum Gravity and Newton's Law}

To YFS resum the propagators in the theory, in
the YFS formula in Eq.(5.16) in Ref.~\cite{yfs}, 
we make the replacements described in Refs.~\cite{bw1,sw2} to
go over from QED to QG and get the factor $e^{B''_g(k)}$ in numerator
of each propagator in Feynman's series~\cite{f1,f2}, 
with $B''_g(k) = \frac{\kappa^2|k^2|}{8\pi^2}\ln\left(\frac{m^2}{m^2+|k^2|}\right)$ in the deep Euclidean regime. If $m$ vanishes, using the usual $-\mu^2$ normalization 
point we get
$B''_g(k)=\frac{\kappa^2|k^2|}{8\pi^2}
\ln\left(\frac{\mu^2}{|k^2|}\right)$. In both cases the
respective resummed propagator falls faster than any power of $|k^2|$!
%\begin{equation}
%i\Delta'_F(k)|_{Resummed} =  \frac{ie^{B''_g(k)}}{(k^2-m^2-\Sigma'_s+i\epsilon)}.
%\label{resum}
%\end{equation}
%This is the basic result.
%Note that
%$\Sigma'_s$ starts in ${\cal O}(\kappa^2)$, so we may drop it in
%calculating one-loop effects. 
%Further,
%explicit evaluation gives, for the deep UV regime,
%\begin{equation}
%B''_g(k) = \frac{\kappa^2|k^2|}{8\pi^2}\ln\left(\frac{m^2}{m^2+|k^2|}\right).
%\label{deep}
%\end{equation}
%The resummed propagator falls faster than any power of $|k^2|$!
%If $m$ vanishes, using the usual $-\mu^2$ normalization 
%point we get
%$B''_g(k)=\frac{\kappa^2|k^2|}{8\pi^2}
%\ln\left(\frac{\mu^2}{|k^2|}\right)$
%which again vanishes faster than any power of $|k^2|$! 
This means that one-loop corrections are finite! 
All quantum gravity loops are UV finite and the all orders
proof
%, as well as the explicit finiteness of ${\Sigma'}$
%at one-loop, 
is given in Refs.~\cite{bw1}.

The one-loop corrections to Newton's law
implied by the diagrams in Fig.~\ref{fig1}
directly impact our black hole issue.
Using the YFS resummed propagators in Fig.~\ref{fig1}  we get
the potential~\cite{bw1,bw2}
$\Phi_{N}(r)= -\frac{G_N M_1M_2}{r}(1-e^{-ar})$
%%\end{equation}
%where $a=1/\sqrt{-\frac{1}{2}\Sigma^{T(2)}}\simeq 3.96 M_{Pl}$
where~\cite{bw1,bw2} $a\simeq 3.96 M_{Pl}$
when for definiteness we set $m\cong 120$GeV.
%We note that
%%\begin{equation}
%$c_2 \cong \ln\frac{1}{\lambda_c}-\ln\ln\frac{1}{\lambda_c}-\frac{\ln\ln\frac{1}{\lambda_c}}{\ln\frac{1}{\lambda_c}-\ln\ln\frac{1}{\lambda_c}}-\frac{11}{6}$.
%%\label{anal1}
%%\end{equation}
%%and we used this result to check our numerical result for $c_2$.
%Without resummation, $\lambda_c=0$
%and $c_2$ is infinite.
%% and, as this is
%%the coefficient of $q^4$ in the inverse propagator, 
%%{\bf no renormalization of the field and/or of the mass could remove
%%such an infinity}. In our new approach, 
%%this infinity is absent.\par
%Our gauge invariant result for $\Sigma^{T(2)}$
%% in (\ref{sigma})
Our gauge invariant analysis
can be shown~\cite{bw1} to be consistent
with the one-loop analysis of QG in Ref.~\cite{thvelt1}\footnote{Our deep Euclidean studies are complementary 
to the low energy studies
of Ref.~\cite{dono1}.}.
%%CROSS CHECK WITH 't HOOFT AND VELTMAN, {\Color{Red}Ann. Inst. Henri Poincare {\bf XX}, 69 (1974)}, {\Color{Black}WHERE THE COMPLETE
%%RESULT OF THE ONE-LOOP DIVERGENCES OF OUR SCALAR FIELD COUPLED
%%TO EINSTEIN'S GRAVITY HAVE BEEN COMPUTED}. 
%%\newpage

%%Sub-Planck scale physics is accessible to point particle field theory
%%so that current superstring theories may be
%%phenomenological models
%%for a more fundamental theory (TUT=The Ultimate Theory) just as the old string theory~\cite{schw1} is such a model for QCD. Other types of correspondences
%%are not excluded here~\cite{superichep04}.
%%Our deep Euclidean studies are complementary 
%%to the low energy studies
%%of Ref.~\cite{dono1}. 
%%The effective cut-off which we generate dynamically
%%is at $M_{Pl}$ so that renormalizable quantum field theory (QFT)  
%%below $M_{Pl}$
%%is unaffected. Some non-renormalizable QFT's are given new 
%%life here -- they may have other problems, however.

\section{Massive Elementary Particles and Black Holes}

%In the SM, there are 
%now believed to be three massive neutrinos~\cite{neut},
%with masses that we estimate at $\sim 3$ eV, and there are 
%the remaining members
%of the known three generations of Dirac fermions 
%$\{e,\mu,\tau,u,d,s,c,b,t\}$. 
With reasonable estimates and measurements 
~\cite{neut,pdg2002,bw1} of the SM particle 
masses, including the various bosons,
%$m_e\cong 0.51$ MeV, $m_\mu \cong 0.106$ GeV, $m_\tau \cong 1.78$ GeV,
%$m_u \cong 5.1$ MeV, $m_d \cong 8.9$ MeV, $m_s \cong 0.17$ GeV,
%$m_c \cong 1.3$ GeV, $m_b \cong 4.5$ GeV and $m_t \cong 174$ GeV,
%as well as the massive vector bosons $W^{\pm},~z$, with masses
%$M_W\cong 80.4$ GeV,~$M_Z\cong 91.19$ GeV.  
%
%To get a better estimate of the size of $c_2$ we 
%use the general spin independence
%of the graviton coupling to matter at {\it relatively low} momentum transfers.
%We count each Dirac fermion as 4 degrees of freedom,
%each massive vector boson as 3 degrees of freedom and remember that
%each quark has three colors. 
the corresponding results for the analogs of the
diagrams in Fig. 1 imply~\cite{bw1}
%$c_2$ in Ref.~\cite{bw2} for each
%SM massive degree of freedom implies approximately
%\begin{equation}
%$c_{2,eff} \cong 9.26\times 10^3$
%\label{ceff} 
%\end{equation}
%so 
that in the SM
%\begin{equation}
$a_{eff} \cong 0.349 M_{Pl}$ .
%\label{aeff} 
%\end{equation}
To make direct contact with black hole physics, 
note that, if $r_S$ is the Schwarzschild radius,
for $r\rightarrow r_S$, $a_{eff}r \ll 1$ so 
that $|2\Phi_{N}(r)|_{m_1=m}/m_2|\ll 1$. This means that
$g_{00}\cong 1+2\Phi_{N}(r)|_{m_1=m}/m_2$ remains 
positive as we pass through the
Schwarzschild radius. 
It can be shown~\cite{bw1} that this 
positivity holds to $r=0$. Similarly, $g_{rr}$ remains negative
through $r_S$ down to $r=0$~\cite{bw1}. 
%To get these results,
%{\color{black}note that in the relevant regime for r, the smallness of
%the quantum corrected newton potential means that we can use the
%linearized einstein equations for a small spherically symmetric
%static source $\rho(r)$ which generates $\phi_{newton}(r)|_{m_1=m}/m_2$
%via the standard poisson's equation.} 
%the usual result {\color{red}(see mtw, abs)} for the
%respective metric solution then gives 
%{\color{pinegreen}$g_{00}\cong 1+2\phi_{newton}(r)|_{m_1=m}/m_2$ and
%$g_{rr}\cong -1+2\phi_{newton}(r)|_{m_1=m}/m_2$ }
%which remain
%respectively time-like and space-like
%to $r=0$.\par
In resummed QG, a massive point particle is not a black hole.\par
%The value of $a_{eff}$ given here is incomplete,
%as there may be as yet unknown massive particles 
%beyond those already
%discovered -- these would only decrease $a_{eff}$.
%Including more particles in the computation of
%$a_{eff}$ would make it smaller and hence would not change the
%conclusions of our analysis. 
%For example, in the minimal 
%supersymmetric Standard
%Model we expect approximately that 
%$a_{eff}\rightarrow \frac{1}{\sqrt{2}}a_{eff}$.

%One can also use the results for the
%complete one-loop UV divergent corrections of Ref.~\cite{thvelt1}
%to see that the remaining 
%interactions at one-loop order not discussed here
%(vertex corrections, pure gravity self-energy corrections, etc. )
%also do not increase the value of $a_{eff}$ --
%$a_{eff}$
%is a parameter which is bounded from above by the estimates we give
%here and which should be determined from cosmological and/or other
%considerations. Such implications will be taken up elsewhere.\par

Our results imply the running Newton constant
$G_N(k)=G_N/(1+\frac{k^2}{a_{eff}^2})$
which is 
fixed point behavior for 
$k^2\rightarrow \infty$,
in agreement with the phenomenological asymptotic safety approach of
Ref.~\cite{reuter2}.
Our result that an elementary particle has no horizon
also agrees with the result in Ref.~\cite{reuter2} that a black hole
with a mass less than
 $M_{cr}\sim M_{Pl}$
has no horizon. The basic physics is the same: $G_N(k)$ vanishes for $k^2\rightarrow \infty$.

Because our value of the coefficient 
of $k^2$ in the denominator of $G_N(k)$
agrees with that found by Ref.~\cite{reuter2}, 
if we use their prescription for the
relationship between $k$ and $r$
in the regime where the lapse function
vanishes,
we get the same Hawking radiation phenomenology as
they do: a very massive black hole evaporates until it reaches a mass
$M_{cr}\sim M_{pl}$
at which the Bekenstein-Hawking temperature vanishes, 
leaving a Planck
scale remnant.

%YFS resummation renders quantum gravity finite
%so that quantum loop corrections are now cut off dynamically.
%Physics below the Planck scale accessible to point particle
%quantum field theory.
%Early universe studies may be able to test predictions.
%  \item
%    renormalizable qft below $m_{pl}$ unaffected
%  \item
% {\color{green} new life to {\color{red}some} nonrenormalizable qft's: {\color{red} they may have other problems}
%We have achieved a minimal union of Bohr's and Einstein's ideas.
%As first checks, we have shown that we are consistent with
%the result in Ref.~\cite{thvelt1} on the one-loop structure
%of quantum gravity and that, contrary to classical expectations,
%a massive elementary SM particle is not a black hole in resummed quantum gravity. Our results are also consistent with the asymptotic safety analysis
%in Ref.~\cite{reuter2} that a black hole of mass less than a critical
%mass $\sim M_{Pl}$ does not have a horizon in quantum gravity
%and that the final state of the Hawking radiation of a massive black hole
%is a Planck scale remnant. Further checks are under investigation.

\section*{Acknowledgments}
We thank Prof. S. Jadach for useful discussions.
This work was partly supported by the US Department of Energy Contract  
DE-FG05-91ER40627
and by NATO Grants PST.CLG.977751,980342.

\end{document}